\documentclass[]{aa}
\usepackage{natbib}
\usepackage{graphicx}
\usepackage{txfonts}

\bibpunct{(}{)}{;}{a}{}{,}

\begin{document}

\title{The effect of MHD turbulence on massive protoplanetary disk fragmentation}

\author{ S\'ebastien Fromang \inst{1,2} }

\offprints{S.Fromang}

\institute{Astronomy Unit, Queen Mary, University of London, 
Mile End Road, London E1 4NS \and Institut d'Astrophysique de Paris, 98Bis Bd Arago,
75014
Paris, France  \\ \email{s.fromang@qmul.ac.uk}}

\newcommand\bb[1] {   \mbox{\boldmath{$#1$}}  }
\newcommand\del{\bb{\nabla}}
\newcommand\bcdot{\bb{\cdot}}
\newcommand\btimes{\bb{\times}}

\date{Accepted; Received; in original form;}

\label{firstpage}

\abstract{Massive disk fragmentation has been suggested to be one of the 
mechanisms leading to the formation of giant planets. While it has been 
heavily studied in quiescent hydrodynamic disks, the effect of MHD 
turbulence arising from the magnetorotational instability (MRI) has never 
been investigated. This paper fills this gap and presents 3D numerical 
simulations of the evolution of locally isothermal, massive and 
magnetized disks. In the absence of magnetic fields, a laminar disk 
fragments and clumps are formed due to the effect of self--gravity. Although 
they disapear in less than a dynamical timescale in the simulations because of 
the limited numerical resolution, various diagnostics suggest that they should 
survive and form giant planets in real disks. When the disk is magnetized, it 
becomes turbulent at the same time as gravitational instabilities develop. 
At intermediate resolution, no fragmentation is observed in these turbulent 
models, while a large number of fragments appear in the equivalent 
hydrodynamical runs. This is because MHD turbulence reduces the strength of 
the gravitational instability. As the resolution is increased, the most 
unstable wavelengths of the MRI are better resolved and small scale 
angular momentum transport starts to play a role: fragments are found to form 
in massive and turbulent disks in that case. All of these results indicate 
that there is a complicated interaction between gravitational instabilities 
and MHD turbulence that influences disk fragmentation processes.
}

\authorrunning{Fromang,S.}
\titlerunning{The effect of MHD turbulence on disk fragmentation}
\maketitle

\section{Introduction}

The recent discovery of more than $100$ extrasolar planets has renewed 
interest in planet formation theory. Because of selection effects, all of 
the planet discovered so far are planets with masses larger than the 
mass of Neptune. However, the problem of their formation remains 
unsolved, despite the attention it continues to enjoy in the recent 
literature \citep{boss98,mayer02,rice03}. Two very different models have 
been proposed. 

In the so-called ``core accretion model,'' giant planets first form their 
cores by dust accumulation. When the core reaches a mass of the order of $10$ 
earth masses, a phase of rapid gas accretion ensues and builds 
up the envelope of the planet \citep{pollack96}. The typical timescale to 
form a planet of $1$ Jupiter mass in a realistic disk is of the order 
of $10^7$ 
years. This rather long timescale has raised questions since it is 
similar to the expected lifetime of the disk. However, recent studies 
suggest that the migration of the planet in the disk may overcome this 
difficulty \citep{hourigan&ward84, rice&armitage03, alibertetal04}.

The alternative scenario is that giant planets form directly in massive 
protoplanetary disks by gravitational fragmentation \citep{boss98,boss00}. 
Massive disks are linearly unstable to axisymmetric gravitational 
instabilities when the Toomre $Q$ parameter \citep{toomre64}, defined by

\begin{equation}
Q=\frac{c_s\kappa}{\pi G \Sigma} \, ,
\end{equation}

\noindent
becomes smaller than $1$. In the definition above, $c_s$ is the sound speed, 
$\kappa$ is the epicyclic frequency (see, e.g.,
\citeauthor{binney&tremaine87}~\citeyear{binney&tremaine87}), 
$\Sigma$ is the disk surface density and $G$ is the gravitational 
constant. Gaseous disks are also unstable to non-axisymmetric perturbations 
when $Q \geq 1$. In the later case, spiral arms develop and transport 
angular momentum outward in the disk. If the nonlinear evolution of the 
instability is 
very violent, these spiral arms may fragment to form dense clumps which have 
been interpreted as proto--giant planets. For example, \citet{gammie01} 
found that a disk fragments when its cooling timescale is shorter than 
$\tau_{cool} \sim 3 \Omega^{-1}$. This was confirmed by \citet{rice03} 
using SPH numerical simulations. 
\citet{mayer02,mayer04} also found fragmentation for a variety of disk models 
and equations of state. However, \citet{pickett00b} failed to obtain 
fragmentation in similar thermodynamic conditions. Whether or not massive disks 
fragment remains controversial. 

As described above, the issue of disk fragmentation has been largely 
addressed in the past few years. These studies omit the effect of magnetic 
forces, despite the known effect of even a weak magnetic field on the 
dynamical 
evolution of a disk. Indeed, rotating disks in which the angular velocity 
decreases outward are unstable to the MRI \citep{balbus&hawley91}. 
In the last decade, numerical simulations, both local and global, have shown 
that its nonlinear outcome is MHD turbulence that transports angular 
momentum outward (see \citeauthor{balbus&hawley98} 
\citeyear{balbus&hawley98}, or \citeauthor{balbusaraa03} 
\citeyear{balbusaraa03},  for a review). Recently, 
\citeauthor{fromangetal04a} (\citeyear{fromangetal04a}, 
\citeyear{fromangetal04b}) investigated the evolution of massive, 
magnetized disks, in which the MRI and gravitational instabilities are 
expected to develop simultaneously. The main result of this 
study is that both instabilities are tighly coupled and strongly interact with 
one another, leading to a 
quasi--periodic variation of the mass accretion rate onto the central star. 
However, these results were obtained using an adiabatic equation of state, 
a regime in which no gravitational collapse was found in the disk. The 
latter is more likely to occur in isothermal disks, since the pressure support 
is less important in that case. The main goal of this paper is to investigate 
the influence of MHD turbulence on the evolution of isothermal disks, with an 
emphasis on the question of their fragmentation.

The plan of the paper is as follows. Section~$2$ presents the numerical 
methods used and the initial conditions of the simulations. Section~$3$ 
describes their results, focusing first on purely hydrodynamic cases and 
then on full MHD simulations. These results are finally discussed in 
section~$4$.

\section{Method}

The relevant equations are those of ideal MHD for a self-gravitating fluid, in 
which $\rho$ is the mass density, 
$\bf{v}$ is the velocity, $\bf{B}$ is the magnetic field, $P$ is 
the gas pressure and $\Phi=\Phi_S+\Phi_c$ is the total gravitational 
potential, which has contributions $\Phi_S$ from the disk self--gravity 
and $\Phi_c$ from a central mass:

\begin{eqnarray}
\frac{\partial \rho}{\partial t} + \del \bcdot (\rho {\bf v})  =  0, \\
\rho \left( \frac{\partial {\bf v}}{\partial t} + ( {\bf v} \bcdot \del )
{\bf v} \right)  =  - \del P - \rho \del \Phi + \frac{1}{4
\pi} (\del \btimes {\bf B}) \btimes {\bf B}, \\
\rho \left( \frac{\partial }{\partial t} + {\bf v} \bcdot \del \right)
\frac{\partial {\bf B}}{\partial t}  =  \del \btimes ( {\bf v} \btimes
{\bf B} ), \\
\nabla^2 \Phi_s = 4 \pi G \rho,
\label{MHD equations}
\end{eqnarray}

\noindent
The equation of state is either adiabatic or locally isothermal. In 
the latter case, the temperature in the disk is a function of position, but 
not time:

\begin{equation}
\frac{P}{\rho}=f(r,\phi,z) \, .
\end{equation}

\noindent
Here $(r,\phi,z)$ denotes the standard cylindrical coordinates that will 
be used throughout the paper. When the equation of state is adiabatic, the 
pressure is derived from the internal energy $e$ using 

\begin{equation}
P=(\gamma - 1) e
\end{equation}

\noindent
where $\gamma=5/3$ is the polytropic index. $e$ is calculated using the 
following energy equation:

\begin{equation}
\rho \left( \frac{\partial }{\partial t} + {\bf v} \bcdot \del \right)
\left( \frac{e}{\rho} \right)  =  -P \del \bcdot {\bf v}, \\
\end{equation}

\subsection{Numerical procedure}

To solve the equations described above, I used the code GLOBAL 
\citep{hawley&stone95}. It uses time--explicit Eulerian finite differences 
and is designed to handle standard cylindrical coordinates. The magnetic 
field is evolved using the combined Method of Characteristics and Constrained
Transport algorithm (MOC--CT), which preserves the divergence of the
magnetic field to machine accuracy at all time. In order to reduce the time of 
the simulations, the computational domain is reduced to $[0,\pi]$ in the 
azimuthal direction, as in \citet{fromangetal04b}. Finally, outflow boundary
conditions are applied in the radial and vertical directions, while periodic
boundary conditions are applied in $\phi$. 

The self--gravitating potential $\Phi_S$ is calculated with a Poisson solver 
well adapted to cylindrical geometry \citep{fromangetal04b}. The density 
distribution of the disk is first Fourier transformed in the azimuthal 
direction. For computational reasons, only those Fourier components $\rho_m$ 
whose index $m$ ranges between $0$ and $m_{max}$ are retained at this stage. 
For each of them, the procedure is the following: (1) From $\rho_m$, the 
gravitational potential $\Phi_S^m$ is obtained at the boundary of the 
computational domain using an expansion in Legendre functions 
\citep{cohl&tohline99}. (2) $\Phi_S^m$ is calculated everywhere on the 
computational domain using the Successive Over Relaxation method 
\citep{hirsh88}. Finally, $\Phi_S$ is reconstructed using the 
values of $\Phi_S^m$, with $m \in [0,m_{max}]$.

One has to understand that the value of $m_{max}$ described above is 
crucial for the problem of disk 
fragmentation addressed in this paper. Indeed, the larger the number of 
coefficient included in the calculation of $\Phi_S$, the smaller the 
structures that can be described by the simulation. Of course, this comes 
also with a larger computational time. Nevertheless, in order to 
accurately describe the tiny structures associated with the possible 
fragmentation of the disks, I used $m_{max}=32$, $64$ or $128$ in this work, 
in contrast with the lower value $m_{max}=8$ typically used by 
\citet[b]{fromangetal04b} in adiabatic simulations where only large structures 
appeared.

\subsection{Initial conditions}
\label{condinit}

\begin{table}[t]
\begin{center}
\begin{tabular}{@{}lccccc}
\hline\hline
Model & $m_{max}$ & $\langle \beta \rangle$ & Resolution & EQS & Fragments?
\\
\hline\hline
HDISO2 & $32$ & $+\infty$ & $(128,64,128)$ & iso & No \\
HDISO3 & $64$ & $+\infty$ & $(128,128,128)$ & iso & Yes \\
HDISO4 & $128$ & $+\infty$ & $(128,256,64)$ & iso & Yes \\
\hline
T2$^*$ & $0$ & $8$ & $(128,64,128)$ & adia & No \\
T3$^*$ & $0$ & $8$ & $(128,128,128)$ & adia & No \\
T4$^*$ & $0$ & $8$ & $(128,256,64)$ & adia & No \\
TISO2 & $32$ & $8$ & $(128,64,128)$ & iso & No \\
TISO3 & $64$ & $8$ & $(128,128,128)$ & iso & No \\
TISO4 & $128$ & $8$ & $(128,256,64)$ & iso & Yes \\
\hline\hline
\end{tabular}
\caption{The first column gives the model label. The second and third 
columns respectively give the number $m_{max}$ of Fourier coefficient 
included in the calculation of the gravitational potential and 
$\langle \beta \rangle$, the initial ratio of thermal to magnetic pressure. 
The resolution $(N_r,N_{\phi},N_z)$ of the model is shown in 
column~$4$, while column~$5$ gives the equation of state used (``iso'' 
stands for locally isothermal and ``adia'' for an adiabatic equation 
of state). Finally, the last column indicates whether or not fragments were 
found in the isothermal runs.}
\label{model properties}
\end{center}
\end{table}

The properties of the various runs performed are summarized in 
table~\ref{model properties}. The first column gives the label of the 
models. HD refers to hydrodynamic models and T refers to runs starting with 
an initial toroidal field. Column~2 gives the number $m_{max}$ of Fourier 
coefficient included in the calculation of the gravitational potential. The 
third column gives the initial ratio $\langle \beta \rangle$ of the volume 
averaged thermal pressure to the volume averaged magnetic pressure. Column~4 
gives the number of grid points $(N_r,N_{\phi},N_z)$ of the run and column~5 
describes the equation of state (``iso'' refers to a locally isothermal 
equation of state). Whether or not fragments form in the disk is described 
in the last column.

Some care has to be taken regarding the initial conditions of each of these 
models. Model T2$^*$ was already presented in \citet{fromangetal04b}: assuming 
an adiabatic equation of state, it calculates the evolution of a disk initially 
in equilibrium, with a mass half that of the central mass, threaded by a 
toroidal weak magnetic field for which the ratio of the volume averaged 
thermal pressure to the volume averaged magnetic pressure 
$\langle \beta \rangle$ is initially $8$. The MRI grows because of the 
magnetic field, and the disk eventually 
becomes fully turbulent. No gravitational instability develops 
since $m_{max}=0$ in this model, which is evolved for about $6$ orbits at 
the location of the initial outer edge of the disk. At this stage, it has 
reached a quasi--stationary turbulent state typical of the outcome of global 
simulations of zero mass disks: mass slowly falls onto the central object 
under the action of the Maxwell and the Reynolds stresses, while angular 
momentum is transported in the outer parts of the disks, which spreads 
as a consequence. At the same time, a tenuous, magnetized corona 
builds up above and below the disk main body. 

This quasi--stationary state emerging from model T2$^*$ after $6$ orbits was 
used as the initial condition for model TISO2. To do so, the temperature was 
calculated at the end of model T2$^*$ everywhere on the grid and then kept 
constant for the further evolution of model TISO2, whose equation of state 
is locally isothermal.

In order to compare model TISO2 with an hydrodynamic model whose properties 
are as closely matched as possible, the disk structure resulting from model 
T2$^*$ was 
also used to compute the initial conditions for model HDISO2. The procedure 
used to do so is as follows: the result of model T2$^*$ was first 
azimuthally averaged and the magnetic fields were removed. A small random 
perturbation was then applied to the disk before restarting the simulation 
with a locally isothermal equation of state. Note that this sudden removal 
of the relatively weak magnetic field does not strongly affect the 
disk global structure.

The other simulations were initialized in a similar way. Model T3$^*$, TISO3, 
HDISO3 and T4$^*$, TISO4, HDISO4 are the high resolution 
equivalent to model T2$^*$, TISO2, HDISO2. For models T4$^*$, TISO4, and 
HDISO4, in which $256$ grid cells were used in the azimuthal direction, the 
number of cells in the vertical direction was decreased by a factor of two. 
This is to save computational time. Even with this reduction of $N_z$, model 
TISO4 required more than 1100 hours of CPU time per orbit (measured at the 
initial outer edge of thi disk) on a 
Pentium 3.06GHz Xeon chip. This is mostly because of the large number of 
Fourier components used in that case ($m_{max}=128$ in both HDISO4 and TISO4).

\section{Results}

\subsection{Hydrodynamic simulations}

\subsubsection{Disk fragmentation}

\begin{figure}
\begin{center}
\includegraphics[scale=0.35]{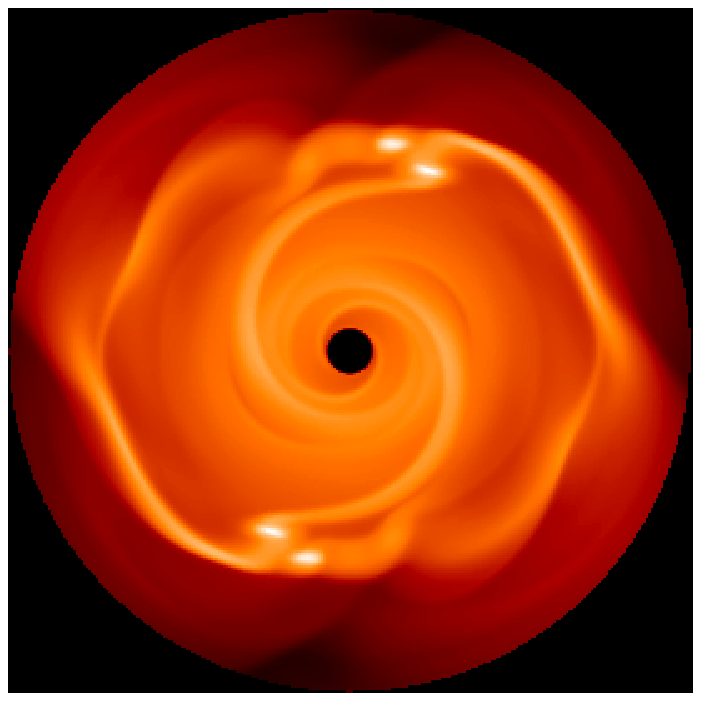}
\includegraphics[scale=0.35]{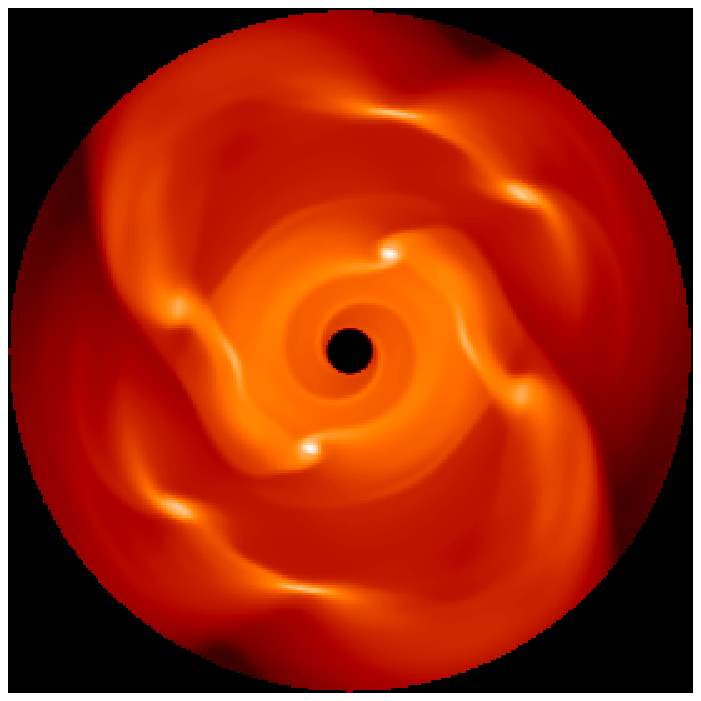}
\includegraphics[scale=0.35]{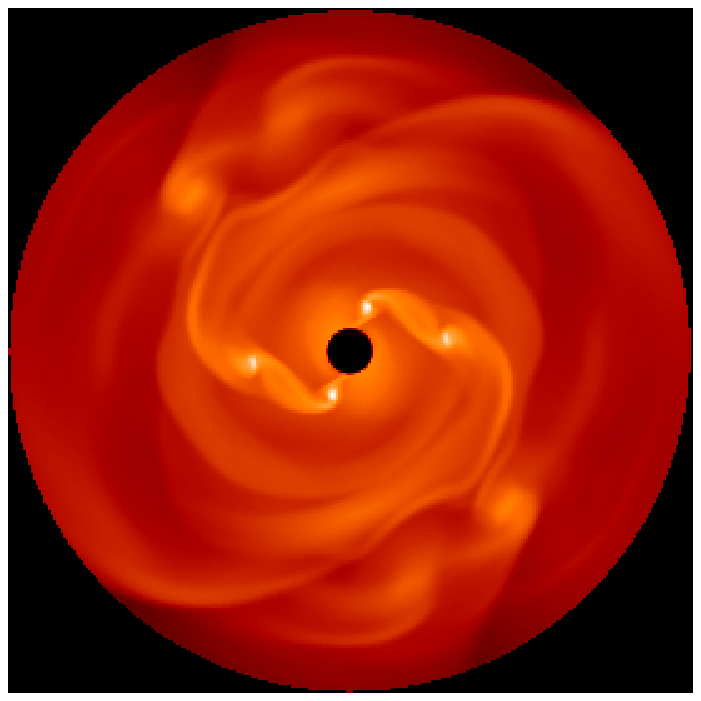}\\
\includegraphics[scale=0.35]{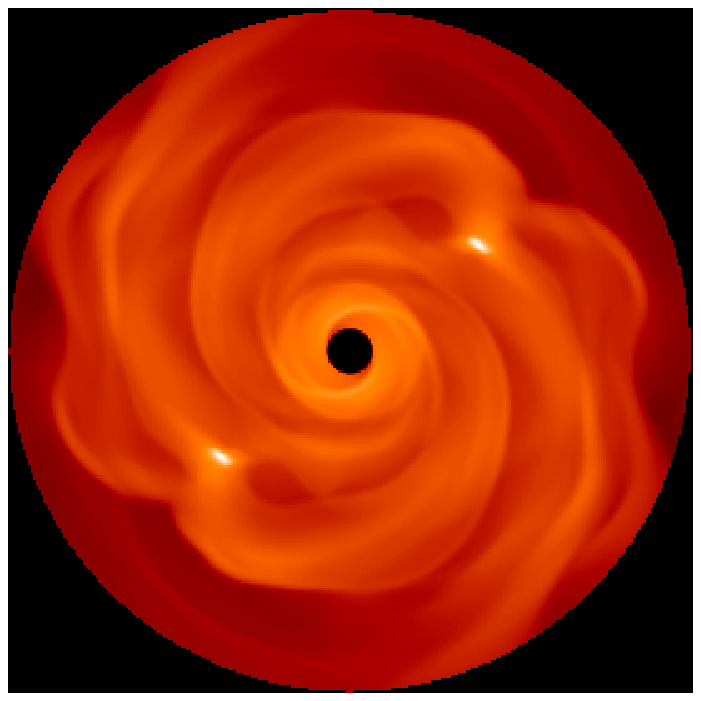}
\includegraphics[scale=0.35]{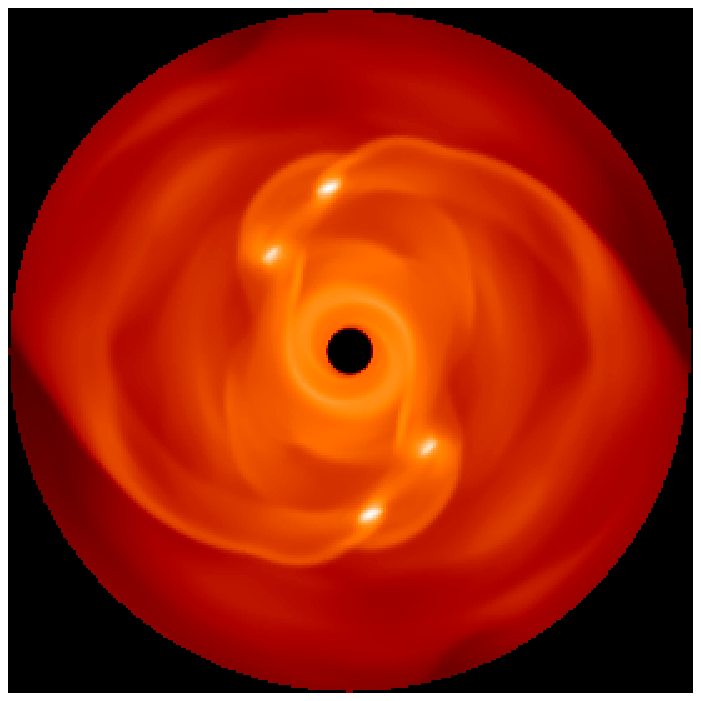}
\includegraphics[scale=0.35]{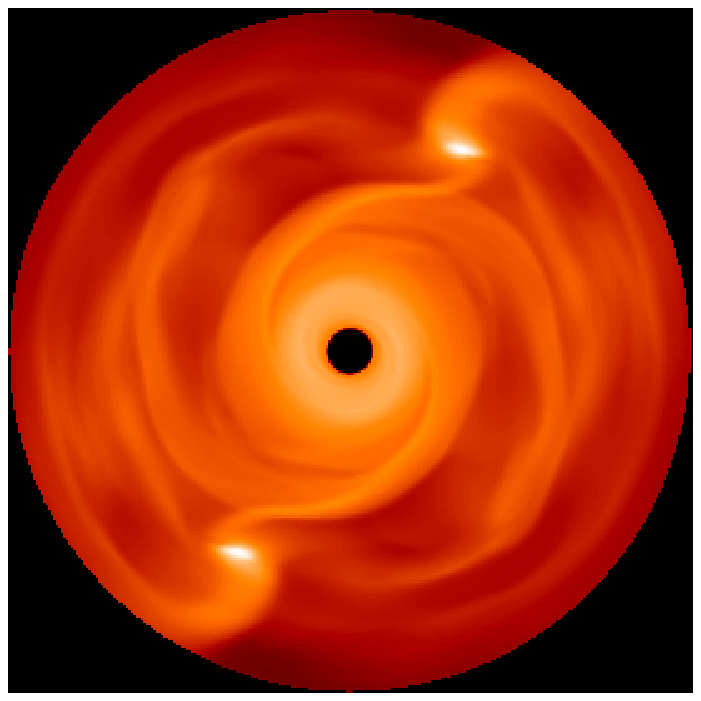}
\caption{Snapshots of the logarithm of the equatorial density of the disk 
for model HD2, respectively at times $t=9.0,9.2,9.3,9.5,9.9$ and $10.3$ 
from top left to bottom right. The structure of the disk varies very 
rapidely. Short lived high density clumps are visible in each snapshots.}
\label{fragment eq}
\end{center}
\end{figure}

I first present the results of the fiducial hydrodynamic model HDISO3, for 
which 
$N_{\phi}=128$. After about one orbit, a strong gravitational instability 
develops in the disk as a consequence of its large mass. The early evolution 
of the instability is similar to what was obtained using an adiabatic 
equation of state \citep{fromangetal04b}. However, the spiral arms that 
form in the present case are stronger because of the weaker pressure support. 
When the gravitational instability saturates, the disk fragments into dense 
clumps. This can be seen in figure~\ref{fragment eq}, which shows six 
snapshots of the logarithm of the density in the equatorial plane of the 
disk at times (measured in the orbital time at the initial outer edge of the 
disk) $9.0$, $9.2$, $9.3$, $9.5$, $9.9$ and $10.3$ (from top 
left to bottom right). Note that the results of the simulation have been 
extended by symmetry to cover the range $[0,2\pi]$ in this figure, which 
explains why an even numbers of clumps is always seen. 
Figure~\ref{fragment eq} clearly illustrates the fact that the state of the 
disk is changing very rapidly. The clumps form, their density increases, 
reaches a maximum and starts to decline. They eventually completely dissolve 
in the background flow in less than the dynamical timescale. They barely 
complete more than a quarter of their orbit before disappearing.

\begin{table}
\begin{center}
\begin{tabular}{@{}lcccc}
\hline\hline
t(orbits) & $9.5$ & $9.9$--$1$ & $9.9$--$2$ & $10.3$ \\
\hline\hline
$a$ & $0.69$ & $0.52$ & $0.65$ & $0.97$ \\
$\rho_{max}$ & $106$ & $357$ & $256$ & $134$ \\
$M_{10}/M_T$ & $3.7 \times 10^{-3}$ & $2.8 \times 10^{-3}$ & $4.4 \times 10^{-3}$ & $2.4 \times 10^{-3}$ \\
$M_{100}/M_T$ & $8.6 \times 10^{-3}$ & $5.8 \times 10^{-3}$ & $7.0 \times 10^{-3}$ & $4.1 \times 10^{-3}$ \\
$M_J/M_T$ & $6.5 \times 10^{-4}$ & $4.3 \times 10^{-4}$ & $4.4 \times 10^{-4}$ & $6.2 \times 10^{-5}$ \\
$r_{clump}$ & $2\times 10^{-2}$ & -- & $2\times 10^{-2}$ & $2\times 10^{-2}$ \\
$r_H$ & $8.3 \times 10^{-2} $ & $5.7 \times 10^{-2} $ & $8.3 \times 10^{-2} $ & $0.1$ \\
$\alpha$ & $0.33$ & $0.5$ & $0.28$ & $0.11$ \\
\hline\hline
\end{tabular}
\caption{Properties of the four clumps visible in the bottom row of 
figure~\ref{fragment eq} (model HDISO3). The first line of the table shows the 
corresponding time at which each of these properties are calculated. It 
serves as a way of labelling the clumps. The following lines then give the 
orbital radius $a$ of the clump, its peak density $\rho_{max}$, its mass 
$M_{10}$ (resp $M_{100}$), ie the mass contained in the volume where the 
density is larger than $\rho_{max}/10$ (resp $\rho_{max}/100$), normalized 
by the total mass of the disk $M_T$, the Jeans mass $M_J$, the radius 
$r_{clump}$ of the clump, its Hill radius and finally the ratio $\alpha$ 
between its thermal and gravitational energy.}
\label{frag properties}
\end{center}
\end{table}

\begin{figure*}
\begin{center}
\includegraphics[scale=0.44]{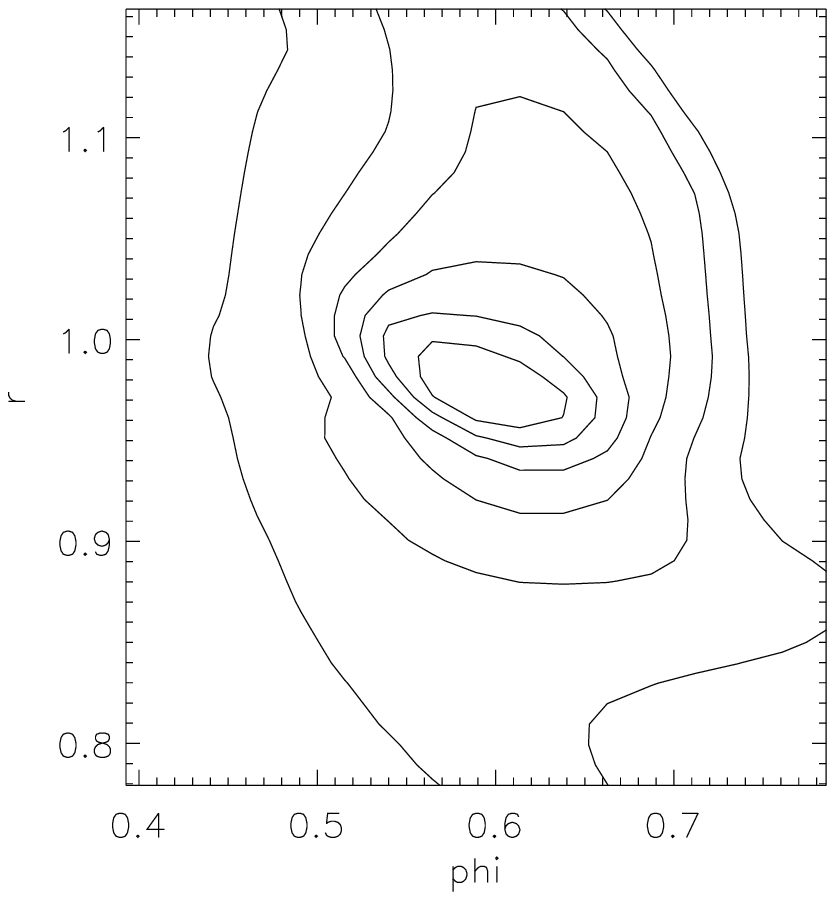}
\includegraphics[scale=0.44]{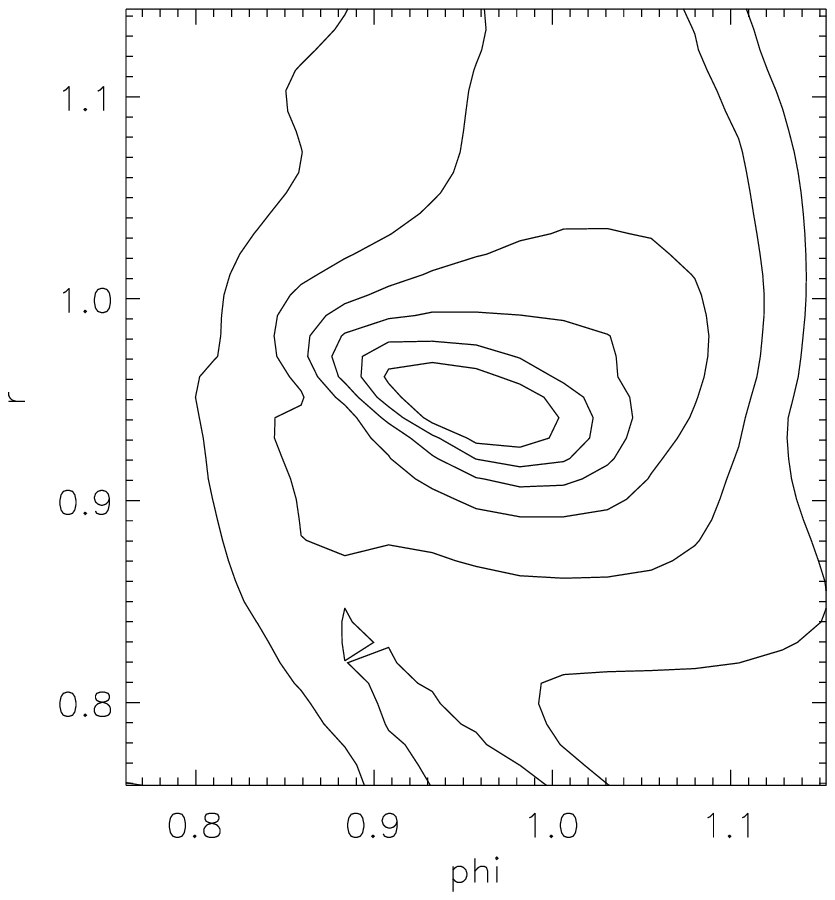}
\includegraphics[scale=0.44]{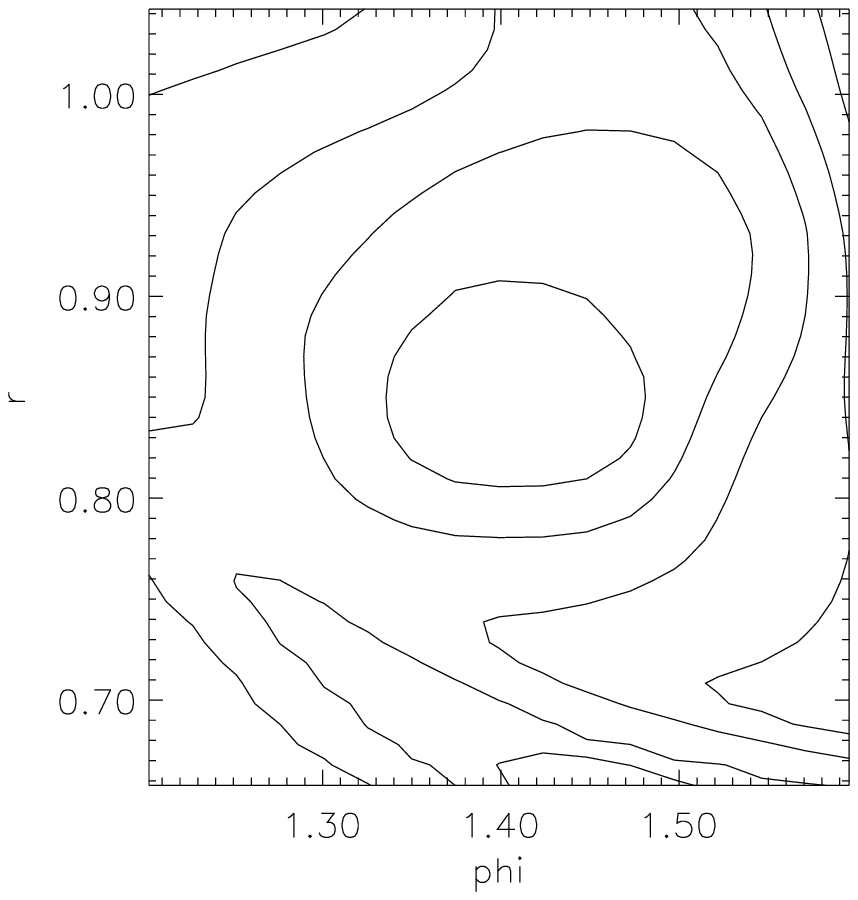}
\caption{Detailed structure of the clump shown in the bottom right frame of 
figure~\ref{fragment eq} at times $t=10.21,10.29$ and $10.37$. $\rho_{max}$ 
is respectively $93$,$70$ and $3.6$ at these times. Six contours levels are 
represented in each panels. They correspond to the density values: $log 
\rho = log \rho_{max}-0.5, log \rho_{max}-1$, etc...}
\label{fragment contour}
\end{center}
\end{figure*}

Table~\ref{frag properties} presents a list of the properties of the 
fragments at times $9.5$, $9.9$ (2 clumps exist in the disk at that time and 
are labelled $9.9$--$1$ and $9.9$--$2$) and $10.3$, when 3D outputs of all the 
variables were saved. The first and second lines show the radius $a$ at 
which each clump has formed and its maximum density $\rho_{max}$. The mass 
$M_{10}$ and $M_{100}$, normalized by the total disk mass $M_T$ are given 
by the third and forth lines. They respectively correspond to the mass of 
the gas contained in the volume where $\rho >\rho_{max}/10$ and 
$\rho >\rho_{max}/100$. This should be compared with the local Jeans mass, 
given 
by the fifth line. Line~$6$ and $7$ respectively give the radius 
of the clump $r_{clump}$ and its Hill radius $r_H$, defined by

\begin{equation}
r_H=a\left(\frac{M_{10}}{3M_c}\right) \, ,
\end{equation}

\noindent
where $M_c=1$ is the central point mass. Finally, the ratio $\alpha$ between 
the thermal and the gravitational energy of the clump is given by the 
last line.

For each cases, the maximum density is above $100$ (the maximum 
density in the initial disk model is $1$). The mass of 
the clumps normalized by the total disk mass is a few times $10^{-3}$. Since 
the latter is of the order of the central point mass, it means that the 
fragments have masses of the order of the mass of Jupiter (for a solar--type 
central mass). The following lines of Table \ref{frag properties} give 
some insights into the stability properties of the clumps: their mass is 
about one order of 
magnitude larger than the local Jeans mass. Their radius is smaller than the 
Hill radius: they are not destroyed by tidal effects. Finally, $\alpha$ 
is smaller than one, showing that they are gravitationally bound.

All these parameters, taken together, indicates that the fragments should 
not disappear. But still, they dissolved very rapidly in the surrounding 
disk. This is detailed in figure~{\ref{fragment contour}}: 
for the clump that formed around $t=10.3$ (bottom right panel of 
figure~\ref{fragment eq}), density contours in the vicinity of the fragment 
are represented. The maximum of the density in the equatorial 
plane of the disk decreases from $93$ to $3.6$ between $t=10.21$ and 
$t=10.37$, during which the fragment performed less than one fifth of its 
orbit. The reason for this disappearance, most likely numerical, is discussed 
in section~\ref{jeans}

\subsubsection{Numerical effects}
\label{num effect}

\begin{figure}
\begin{center}
\includegraphics[scale=.7]{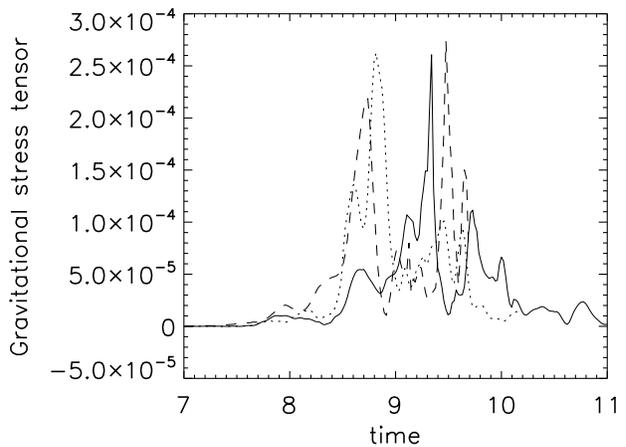}
\caption{Time evolution of the volume averaged gravitational stress tensor 
for model HDISO2 ({\it dotted line}), HDISO3 ({\it solid line}) and 
HDISO4 ({\it dashed line}). The peak value of 
$\langle T_{r\phi}^{grav} \rangle (t)$ is comparable in all models 
even though their resolution in the azimuthal direction is different.}
\label{iso grav stress}
\end{center}
\end{figure}

Given the difficulties in accurately describing fragmentation 
in protoplanetary disks, it is important to investigate the numerical 
artifacts that could influence the outcome of the simulation presented 
above. The most important of such effects is obviously numerical resolution, 
especially in the azimuthal direction. In model HDISO2, $N_{\phi}$ equals 
$64$, while in model HDISO4, $N_{\phi}=256$.

The qualitative evolution of model HDISO4 is very similar to that of 
model HDISO3: fragments form rapidly after the gravitational instability 
saturates. The maximum density they reach in larger ($\rho_{max} \sim 500$), 
but, like model HDISO3, they disappear very quickly. For the low resolution 
run HDISO2, however, no fragment 
appears at all in the disk. This trend shows how sensitive the issue of 
resolution is on disk fragmentation. Indeed, this qualitative difference 
arises even though the development of the instability seems to be very 
similar for all models. This can been seen in figure~\ref{iso grav stress}, 
which shows the time history of the volume averaged gravitational stress 
tensor. It is defined as \citep{fromangetal04b}:

\begin{equation}
\langle T_{r\phi}^{grav} \rangle (t) = \frac{1}{4\pi G}  \int_{V} {
\frac{\partial \Phi_s}{\partial r} \frac{\partial
\Phi_s}{\partial \theta} drd\theta dz} \, .
\end{equation}

\noindent
The time history of $\langle T_{r\phi}^{grav} \rangle$ is shown in 
figure~\ref{iso grav stress} for model HDISO2 
({\it dotted line}), HDISO3 ({\it solid line}) and HDISO4 ({\it dashed line}). 
For each cases, it starts to rise after about $8.5$ orbits, and reaches 
similar 
peak values of $2.5 \times 10^{-4}$. The curves are all very similar, showing 
that the azimuthal resolution has little effect on the strength of the 
instability itself, despite the qualitative differences in the fragmentation 
of the disk.

\subsection{MHD simulations}

\begin{figure}
\begin{center}
\includegraphics[scale=0.7]{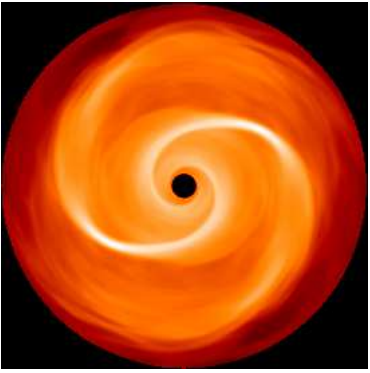}
\caption{Logarithm of the density in the equatorial plane of a turbulent 
disk (model TISO3) at time $t=6.3$, when the density reaches its largest 
value $\rho_{max}=16$.}
\label{image TISO3}
\end{center}
\end{figure}

\begin{figure}
\begin{center}
\includegraphics[scale=0.53]{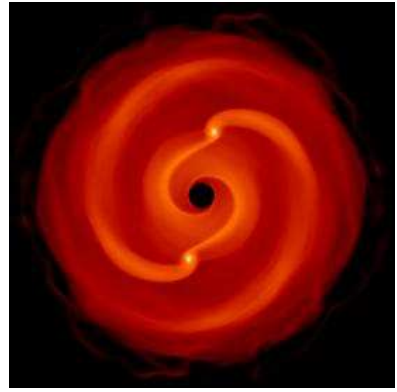}
\caption{Same as figure~\ref{image TISO3}, but for model TISO4 at 
time $t=6.2$. The largest value of the density in this case is 
$\rho_{max}=550$.}
\label{image TISO4}
\end{center}
\end{figure}

In this section, I compare the results of model TISO3 and TISO4 with their 
hydrodynamic counterpart HDISO3 and HDISO4. The results of model TISO2, 
for which $N_{\phi}=64$, are very similar to those of model TISO3 and 
will not be described further. 

In both TISO3 and TISO4, a gravitational instability quickly develops 
on top of MHD turbulence. However, the qualitative evolution of the 
instability is dramatically different in both models. In the lower resolution 
case TISO3, 
no disk fragmentation is observed, while fragments quickly form in the high 
resolution model TISO4. This is illustrated by figure~\ref{image TISO3} 
and \ref{image TISO4}. The former shows the logarithm of the density 
in the midplane of the disk for model TISO3 after $6.2$ orbits, 
which corresponds to the time when it reaches its largest density, 
$\rho_{max} \sim 16$. Note how different the disk structure looks compared 
to its hydrodynamic 
equivalent HDISO3 (figure~\ref{fragment eq}). On the contrary, 
figure~\ref{image TISO4} clearly shows that fragments formed in model 
TISO4, despite the presence of MHD turbulence. The properties and time history 
of the clumps are very similar to those observed in the hydrodynamic models 
described above. The likely origin of the difference between models TISO3 and 
TISO4 is discussed in the following section.

\section{Discussion}

\subsection{The Jeans length}
\label{jeans}

\begin{figure}
\begin{center}
\includegraphics[scale=0.45]{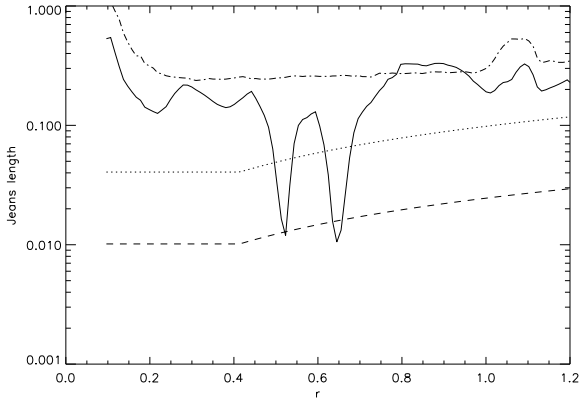}
\caption{Jeans length vs radius for model HDIOS3 at time $t=0$ 
({\it dotted-dashed line}) and $t=9.9$ ({\it solid line}). The dashed line 
shows the maximum of the radial and azimuthal spacing of the grid as a 
function of radius. The dotted line plots four times this value.}
\label{Jeans hydro}
\end{center}
\end{figure}

\begin{figure}
\begin{center}
\includegraphics[scale=0.45]{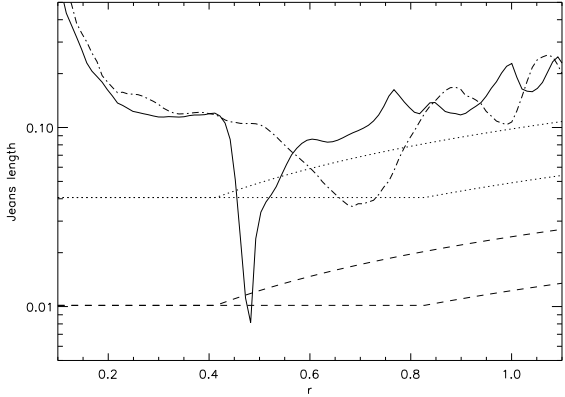}
\caption{Same as figure~\ref{Jeans hydro}, but for models TISO3 
({\it dotted-dashed line}) at time $6.3$ and TISO4 at time $6.2$. The dotted 
and dashed lines have the same meaning as for figure~\ref{Jeans hydro} 
(the upper curves show the case of model TISO3, for which $N_\phi=128$, while 
the lower curves correspond to model TISO4).}
\label{Jeans mhd}
\end{center}
\end{figure}

A critical parameter to take into account in the simulations presented 
above is the Jeans length, defined by:

\begin{equation}
\lambda_J=\left( \frac{\pi c_s^2}{G \rho} \right)^{1/2}
\end{equation}

\noindent
\citet{trueloveetal97} showed that spurious 
fragmentation can occur if the Jeans length is not well resolved at all times 
during the disk evolution and that the largest 
grid cell on the computational domain should be about $4$ times larger than 
$\lambda_J$. In isothermal simulations, it is very difficult to resolve the 
Jeans length during the disk fragmentation: since the sound speed is 
constant, $\lambda_J$ increases as $(1/\rho)^{1/2}$. It is therefore 
important to follow its behaviour. At each radii, one can calculate the 
minimum value of $\lambda_J$ in the equatorial plane. Its radial profile is 
represented in figure~\ref{Jeans hydro} for model HDISO3 at times $t=0$ 
({\it dotted-dashed line}) and $t=9.9$ ({\it solid line}). The dashed line 
represents the largest grid size as a function of $r$ and the dotted line 
shows $4$ times this value. At $t=0$, it is obvious that the Jeans 
length is well resolved 
and satisfies the Truelove criterion. However, at $t=9.9$, two fragments 
formed in the disk and the resolution of the simulation becomes coarser than 
the local Jeans length. According to the Truelove criterion, the 
evolution of the fragments after this point is not meaningful and should 
not be regarded as being physical. On top of this, the size of 
the forming clumps themselves at this stage of the simulation is of the order 
of a few grid cells. The gravitational potential they create (and which 
governs their subsequent collapse) is not accurately evaluated at 
this scale because of the finite resolution of the grid. Their later evolution 
in the simulation is not well described. Both of these arguments show that 
these calculations can only be used as a diagnostic for disk fragmentation. 

As a conclusion, the hydrodynamic simulations tell us that 
massive quiescent disk 
should fragment in the conditions studied here. However, they cannot tell us 
anything about their long term evolution, even though their stability 
properties, presented above, suggest that they may survive.

Figure~\ref{Jeans mhd} is similar to figure~\ref{Jeans hydro}: it 
compares the grid resolution with the local Jeans length for the MHD turbulent 
disk models TISO3 ({\it dotted-dashed line}) and TISO4 ({\it solid line}) 
respectively at times $t=6.2$ and $t=6.3$ (ie at the same time as the 
snapshots shown 
in figure~\ref{image TISO3} and \ref{image TISO4}). In the case of model 
TISO3, the minimum value of the local Jeans length (which decreases at the 
location of the spiral arms), is roughly three times the resolution of the 
simulation. The Truelove criterion is only marginally violated in that 
case. However, in similar conditions, clumps are seen to form in model 
HDISO3, 
which shows that MHD turbulence clearly has an effect in that case. 
Figure~\ref{Jeans mhd} also shows that the case of model TISO4 is very 
similar to the hydrodynamic models for which fragmentation is found: the 
local Jeans length decreases as the density of the fragments increases, until 
the former becomes smaller than the resolution. The subsequent evolution of 
the clump is not accurately described by the simulations. 

\subsection{The effect of MHD turbulence}

\begin{figure}
\begin{center}
\includegraphics[scale=0.7]{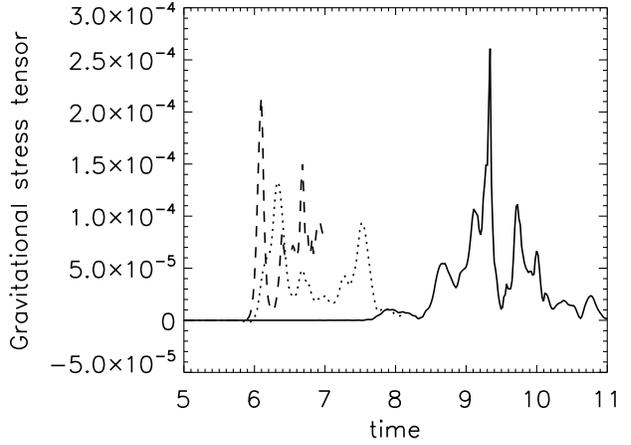}
\caption{Time history of the volume averaged gravitational stress tensor for 
models HDISO3 ({\it solid line}), TISO3 ({\it dotted line}) and 
TISO4 ({\it dashed line}). The earlier growth of the gravitational 
instability in the MHD cases is triggered by the large density perturbations 
due to MHD turbulence. }
\label{compar grav tensor}
\end{center}
\end{figure}

\noindent
The results of model TISO3 and TISO4, together with those of their 
hydrodynamic counterpart HDISO3 and HDISO4 suggest that MHD turbulence 
affects the 
outcome of gravitational instabilities in disks. Although the precise 
interplay between these processes is difficult to understand given the 
complexity of the problem and its sensitivity to numerical resolution, some 
indications are provided by a detailed analysis of the simulations. 
Figure~\ref{compar grav tensor} shows the time 
history of the volume averaged gravitational stress tensor 
$\langle T_{r\phi}^{grav} \rangle$ (defined in section~\ref{num effect}) for 
models HDISO3 ({\it solid line}), TISO3 ({\it dotted line}) and 
TISO4 ({\it dashed line}). The maximum value reached by 
$\langle T_{r\phi}^{grav} \rangle$ is smaller when the disk is turbulent, 
suggesting that the strength of the gravitational instability itself is 
reduced by the presence of the turbulence. It may explain 
why no fragments form in model TISO3, as the gravitational stress is 
decreased by roughly a factor of two compared to model HDISO3 (Note 
that, even though the evolution of the stress is shown for a rather short 
period of time, the difference between the various runs is meaningful because 
this is precisely during that time that fragments forms or not). This effect 
of MHD turbulence on the strength of the gravitational 
instability was already noted in simulations using an adiabatic equation of 
state \citep{fromangetal04b}. In adiabatic simulations, this 
decrease of the gravitational stress did not affect the qualitative 
outcome of the gravitational instability. However, in model TISO3 it prevents 
the density to become large enough in the spiral arm to trigger their 
collapse. 

Figure~\ref{compar grav tensor} also shows that 
$\langle T_{r\phi}^{grav} \rangle$ is larger in model TISO4 (in which clumps 
form) than in TISO3. This is not due to the gravitational instability 
alone, since no such dependency is observed in the hydrodynamic models 
HDISO2, HDISO3 and HDISO4. Nor is it due to the MHD turbulence alone, since 
its properties are similar in model TISO3 and TISO4. However, as is explained 
below, the disk fragmentation (as well as the increase of the gravitational 
stress tensor) observed in model TISO4 
could be due to the small scale angular transport properties of the MRI. In 
cases, such as the present one, when the magnetic field is mostly toroidal, 
the most unstable mode of the MRI has an azimuthal wavenumber $m_{crit}$ that 
satisfies the relation \citep{balbus&hawley98}

\begin{equation}
\frac{m_{crit}}{R} v_A \sim \Omega \, ,
\end{equation}

\noindent
which gives

\begin{equation}
m_{crit} \sim \frac{R\Omega}{c_s}\sqrt{\frac{\beta}{2}} \, .
\end{equation}

\noindent
At the position where the fragment 
forms in model TISO4 (see figure \ref{image TISO4}), the properties of the 
disk model indicate that $R\Omega/c_s \sim 16$, while the turbulence is 
characterized by $\beta \sim 10$. These values gives $m_{crit} \sim 35$. Such a 
mode is barely resolved in model TISO3 (only $7$ grid cells per wavelength) but 
much better described in model TISO4 which has twice the resolution. In 
the latter case, it grows on an orbital timescale and helps to remove the 
angular momentum of the collapsing body. In the former case, the growth 
rate of this small scale mode is lower and it is likely it will be 
less important in the collapse process.

To summarize, these results suggests that, on top of the Jeans length, it is 
important in calculations of the present type to well resolve the length 
$\lambda_B$ defined by

\begin{equation}
\lambda_B = v_a P
\end{equation}

\noindent
where $P$ is the orbital period in the disk.

Finally, it is interesting to compare the present results to 
those of \citet{kimetal03} who performed local calculations of giant molecular 
cloud formation in the presence of MHD turbulence. They 
found that density fluctuations triggered by the MRI can be swing--amplified 
and lead to disk fragmentation. This is not observed here. The reason for this 
difference lie in the initial setup of 
both calculations: \citet{kimetal03} start their simulations with vertical 
fields with a non--zero net flux that produces a stronger turbulence than in 
the present work. Indeed, the ratio of the total stress to the pressure they 
obtain is $\sim 0.2$, while it is an order of magnitude smaller here 
\citep{fromangetal04b}. The larger density fluctuations that result are more 
likely to be gravitationnally bound in their case than in the present 
one.

\section{Conclusion}

This paper presents a collection of hydrodynamic and magnetohydrodynamic 
3D numerical simulations of the evolution of massive isothermal disks. 

In hydrodynamic simulations, the disk is found to fragment into high density 
clumps when the number of grid cells $N_{\phi}$ is larger than $128$ in the 
azimuthal direction. The fragments, whose mass is comparable to the mass 
of Jupiter, disappear very quickly in the disk. However, due to the limited 
resolution of the simulations, the local Jeans length is no longer resolved 
when their density has increased by about two orders of magnitudes. 
Moreover, at these small scales, the collapse of the clumps themselves is 
not accurately described because the gravitational potential is smoothed at 
the grid scale. This is probably what causes their quick disappearance. At 
the very least, the long term behavior of these fragments cannot be studied 
with these simulations, although various criteria suggest that they could 
survive for an extended period of time.

When a weak magnetic field is present initially in the disk, it becomes 
turbulent because of the nonlinear development of the MRI. This turbulent 
nature of the flow has an effect on the evolution of the gravitational 
instability: no fragment is found when $N_{\phi}=128$, in contrast with the 
hydrodynamic model having the same resolution. However, fragments appear 
when the azimuthal resolution is increased to $N_{\phi}=256$. An analysis of 
these results suggest that they are a consequence of two competing effects: on 
the one hand, MHD turbulence tends to decrease the strength of the 
gravitational 
instability, as was already noted in a previous study \citep{fromangetal04b}: 
this is sufficient to suppress gravitational fragmentation in the lower 
resolution case. On the other hand, when the most unstable modes of the MRI 
are resolved, small scale angular momentum transport appears to help the 
formation of bound fragments. This interplay between disk fragmentation 
and radial transfer of angular momentum within an overdense region was 
already suggested by \citet{kimetal03}.

Despite the difficulties related to the limited numerical resolutions of the 
simulations, the results presented in this paper indicate that 
MHD turbulence is an important component in the process of massive disk 
fragmentation. Future studies are needed to quantify the importance 
of the turbulence and to investigate whether the fragments that form are 
truly long lived or quickly dissolve in the background flow. To cope with 
the large resolutions needed to adress this problem, adaptative mesh 
refinement techniques, possibly in cylindrical coordinates, may prove to be 
the method of choice.

\section*{ACKNOWLEDGMENTS}
The simulations presented in this paper were performed at
the Institut du D\'eveloppement et des Resources en Informatique
Scientifique and at Queen Mary University of London.

\noindent
The author acknowledges C.Terquem and S.Balbus whose comments on 
an earlier draft of this paper greatly improved its contents.

\bibliographystyle{aa}
\bibliography{author}

\end{document}